\documentclass[amsmath,amssymb,
preprint
]{revtex4}
\usepackage{graphicx}
\usepackage{dcolumn}
\usepackage{bm}

\begin{document}

\title{The best nanoparticle size distribution for minimum thermal conductivity}

\pacs{ }


\author{Hang Zhang}
\affiliation{Division of Engineering and Applied Science\\
California Institute of Technology\\
Pasadena, CA 91125}

\author{Austin J. Minnich}
\email{aminnich@caltech.edu}
\affiliation{Division of Engineering and Applied Science\\
California Institute of Technology\\
Pasadena, CA 91125}

\begin{abstract}

Which sizes of nanoparticles embedded in a crystalline solid yield the lowest thermal conductivity? Nanoparticles have long been demonstrated to reduce the thermal conductivity of crystals by scattering phonons, but most previous works assumed the nanoparticles to have a single size. Here, we use optimization methods to show that the best nanoparticle size distribution to scatter the broad thermal phonon spectrum is not a similarly broad distribution but rather several discrete peaks at well-chosen nanoparticle radii. For SiGe, the best size distribution yields a thermal conductivity below that of amorphous silicon. Further, we demonstrate that a simplified distribution yields nearly the same low thermal conductivity and can be readily fabricated. Our work provides important insights into how to manipulate the full spectrum of phonons and will guide the design of more efficient thermoelectric materials.
 
\end{abstract}

\maketitle

Engineering the thermal conductivity of solids is important for many applications, ranging from thermal management of electronic devices to thermoelectric energy conversion\cite{Dresselhaus2007,eric_pop2010,Cahill2003,Cahill2014}. In recent years, numerous works have demonstrated that nanostructured materials such as superlattices\cite{segregation2013,Venkatasubramanian2009,Cahill_superlattice_1997}, crystals with embedded nanoparticles\cite{Kim2006,Mingo2009,Pernot2010}, nanocomposites\cite{Poudel02052008,Mehta2012,Biswas2011}, and all-scale hierarchical structures\cite{hierarchical2012} can substantially reduce thermal conductivity of solids, even below the alloy limit\cite{segregation2013}. Many of these materials are under consideration as thermoelectric materials\cite{Kim2006,hierarchical2012,Poudel02052008,Mehta2012,Biswas2011}. Theoretical studies have attributed these reductions in thermal conductivities to scattering of phonons from interfaces and boundaries\cite{Bera2010,Jeng2008,Chen1998}.

A number of studies have examined the effect of nanoparticles with a different mass than that of the host lattice on the thermal conductivity of the composite\cite{Mingo2009, Kim2006, Majumdar1993,Kundu2011}. Majumdar predicted the thermal conductivity of a composite using a model for nanoparticle scattering that interpolates between the Rayleigh scattering and geometrical scattering limits\cite{Majumdar1993}. Recently, advances in first-principles calculations enable the scattering rate from nanoparticles to be calculated without any adjustable parameters\cite{Kundu2011}. Experimentally, very low thermal conductivities below the alloy limit were reported in InGaAs alloys with ErAs embedded nanoparticles \cite{Kim2006}. Mingo et al. introduced a nanoparticle-in-alloy thermoelectric concept that was predicted to have high thermoelectric figure of merit\cite{Mingo2009}. 

Despite extensive study of the thermal properties of solids with nanoparticles, most works did not consider the size distribution of the nanoparticles. The distribution is expected to be important as a number of recent experimental and theoretical works have demonstrated that thermal phonons possess a very broad mean free path (MFP) spectrum\cite{ Esfarjani2011,Henry2008}, making it unlikely that a single nanoparticle size can effectively scatter the entire spectrum. This hypothesis is supported by a recent study of SiGe superlattices, which demonstrated that alloy limit can be broken by a superlattice that combines point defects and abrupt boundaries\cite{segregation2013}. Such a structure is able to achieve a lower thermal conductivity than either a pure alloy or a pure superlattice by scattering both high and low frequency phonons by point defects and abrupt boundaries, respectively.

This observation naturally leads to a simple question regarding crystals containing nanoparticles. Supposing that we can choose any distribution of nanoparticle sizes, which one should we choose to obtain the minimum thermal conductivity? Here, we use numerical optimization methods to answer this question. We find that the broad spectrum of thermal phonons can be most effectively scattered not by a similarly broad distribution but by one with a few discrete peaks at well-chosen radii. These optimized structures achieve even lower thermal conductivity than previously reported nanoparticle-in-alloy structures of Mingo et al. \cite{Mingo2009}. Our result enables a better understanding of how to achieve the minimum thermal conductivity in crystals and will guide the development of more efficient thermoelectrics.

\section*{Modeling}
We begin by considering the thermal conductivity of a crystalline solid, which we take to be Si, with a fixed volumetric fraction of Ge mass defects $x$ in the lattice. The thermal conductivity $k$ of an isotropic crystal is given by:
\begin{equation} \label{eqn}
k= \int \frac{1}{3}C(\omega)v^2(\omega)\tau(\omega)d\omega
\end{equation}
where $C(\omega)=\hbar\omega D(\omega)\partial f/\partial T$ is the spectral specific heat of phonons and $\hbar$ is reduced Plank constant, $D(\omega)$ is the density of states, $f$ is the Bose-Einstein distribution, $T$ is temperature, $v$ is the group velocity, and $\tau$ is the relaxation time. We assume that phonons are scattered by phonon-phonon interactions as well as variable diameter nanoparticles that are formed from the available mass defects. Based on Mathiessen's rule, the total relaxation time can be expressed as $\tau^{-1}=\tau_a^{-1}+\tau_{np}^{-1}$, where $\tau_a$ and $\tau_{np}$ are the relaxation times for anharmonic and nanoparticle scattering, respectively. We consider several model dispersions in this work, including a Debye model in [100] direction, a Born von Karman (BvK) model in [100] direction and the full dispersion for Si calculated by density functional theory (DFT) with all the phonon modes. The dispersion and relaxation time constants for the Debye model and BvK model are taken from ref \citenum{Wang2011}, while those for DFT were calculated by Jes\'us Carrete and N. Mingo using ShengBTE \cite{Li2014,shengbte} and Phonopy\cite{phonopy}, from interatomic force constants obtained with VASP\cite{ Kresse1993, Kresse1994, Kresse1996, Kresse1996a}. The thermal conductivity for pure Si for the DFT dispersion is 166 W/mK while that for the other two dispersions is approximately 150 W/mK at room temperature.

By treating Ge atoms as independent point-defects in a silicon matrix, Garg et al successfully reproduced the experimental thermal conductivities of SiGe alloys, even at Ge concentrations of up to 50\%\cite{Garg2011}. Following this work, we assume that each nanoparticle can be treated independently. Recent works have demonstrated that scattering rates from nanoparticles can be predicted without any adjustable parameters using DFT and atomistic Green's functions\cite{Kundu2011}, but at significant computational cost. To simplify the computations, here we use a previously reported model for nanoparticle scattering rates that interpolates between the short and long wavelength regimes:\cite{Majumdar1993,Mingo2009}
\begin{equation} 
\tau_{np}^{-1}=v(\sigma_s^{-1}+\sigma_l^{-1})^{-1}n
\end{equation}
where, for short and long wavelength regimes we have: $\sigma_s=2\pi R^2$ and $\sigma_l=\frac{4}{9}\pi R^2(\Delta\rho/\rho)^2(\omega R/v)^4$, respectively, where, $ \Delta\rho $ is the density difference between embedded nanoparticles and the bulk material; and $\rho$ is the density of the bulk material. Despite its simplicity, this model has been shown to be match the exact scattering calculations to within around 20\% \cite{Kundu2011}. The available Ge defects can then be distributed into nanoparticles of variable radius $ R $. The smallest nanoparticle is a point defect, which is a single Ge atom in Si lattice. This model neglects differences in interatomic force constants between Si and SiGe. However, prior first-principles work has shown that the dominant source of phonon scattering in alloys is the mass difference, and thus changes in force constants can be neglected without affecting our conclusions\cite{Kundu2011}. This model predicts a thermal conductivity of 10 W/mK for $Si_{0.5}Ge_{0.5}$ alloy, which agrees with the experimental result\cite{ Maycock1967}. 

We can now mathematically pose the problem of minimizing the thermal conductivity. We seek to identify the nanoparticle size distribution, or the fraction of mass defects allocated to nanoparticles of different sizes, that minimizes Eq. 1. This minimization is subject to the constraint that the total volumetric fraction of mass defects $x$ is fixed:  
\begin{equation} \label{example}
V_{per}=\sum_ix(R_i)=\sum_i\frac{4}{3}\pi R_i^3n_i
\end{equation}
where $ R_i $ is the radius of the nanoparticles and $n_i$ is the number density of nanoparticles of radius $ R_i$.

This optimization is difficult to perform using traditional methods such as convex optimization due to the nonlinearity of the function to be minimized and the large number of variables. To overcome this challenge, we use a Particle Swarm Optimization algorithm, which is a recently developed evolutionary algorithm \cite{Kennedy1995,Shi1998}. The algorithm optimizes a multi-dimensional function by iteratively improving candidate solutions according to a fitness function. Each particle's movement is guided by its local best solution as well as the global best solution discovered by other particles. Thus, these particles iteratively approach the best solution in the search-space. Since searching operations tend to cluster in regions of search space with the best solution, the algorithm is much more efficient than exhaustively searching all possibilities, which in the present case is impossible due to the enormous number of possible solutions. 

To implement the algorithm, we first discretize the nanoparticle radii into $M$ = 40 bins ranging from 0.136 to 5.44 nm in increments of 0.136 nm; an isolated point defect corresponds to a Ge atom in crystal Si with an effective radius of 0.136 nm, while a 0.27 nm particle consists of 8 atoms, and so on. Larger size ranges up to $R$=50 nm have also been studied. However, very similar or exactly the same optimal size distributions and thermal conductivities were obtained from these larger ranges. To avoid considering the potential complicated phonon transport inside large nanoparticles, we set a threshold for the largest particle size at $R$=5.44 nm. We divide the total volumetric percentage of mass defects into $N\sim~$100-2000 shares, and initialize the system by randomly allocating the $N$ shares among the $M$ bins. Therefore, each candidate solution can be considered as a vector in an $N$-dimensional search-space describing the number of shares in each bin. Note that each share corresponds to a fixed volumetric percentage of Ge defects and therefore represents different number density of nanoparticles depending on the nanoparticle radius.

The algorithm starts by moving the first share to all other bins and identifying the bin that achieves the lowest thermal conductivity. The share is placed in this bin, and the same procedure is performed for the second share. This procedure repeats until all shares have been moved. Then, the algorithm restarts the cycle and performs the same operations. The program continues this procedure until the thermal conductivity does not decrease further. Usually the optimized distribution can be achieved in 20 cycles, and therefore the computational expense is only of the order $20\times M\times N$. 

\section*{Results}
We present the best size distribution that minimizes the thermal conductivity as a volumetric percent versus nanoparticle radius in Figure 1. We performed the optimization for the Debye, BvK, and DFT dispersion with a Ge volume fraction of 1\%. Independent of the initial starting guess and the number of bins, for the distribution or the dispersion, the algorithm always selects a similar size distribution of three discrete peaks. No other sizes of nanoparticles exist in these configurations. We tested a variety of initial conditions, number of bins, and number of shares, and verified that the algorithm always chooses the essentially the same configuration: the same number of peaks, same peak positions and nearly the same Ge fraction in each peak. What is more, thermal conductivities yielded from these optimal configurations are the same to within 0.5\%. We also verified the algorithm's result for small $N$ by calculating the thermal conductivity for all possible combinations and confirming the chosen configuration is the global minimum. Therefore, the algorithm is choosing the best configuration that minimizes the thermal conductivity among all possible configurations. 

Figure 1 demonstrates that the distribution that most effectively scatters the broad phonon spectrum is actually a series of discrete peaks.  This result is somewhat unexpected, because one might expect that to scatter a broad spectrum of phonons one should also choose a broad distribution of nanoparticle sizes. However, the distribution in Figure 1 results in a substantially lower thermal conductivity than that corresponding $Si_{0.99}Ge_{0.01}$ alloy counterpart. For example in Figure 1c, using the DFT dispersion, we calculate that the thermal conductivity corresponding to an alloy configuration is $k_{alloy}$ = 86 W/mK, which is consistent with previously reported experimental value\cite{ Maycock1967}; while the optimized distribution gives a thermal conductivity of $k_{opt}$ = 30 W/mK.

To gain further insight, we calculate the best distribution for several different volumetric percentages of Ge, shown in Figure 2. We observe additional peaks appear in the optimal distribution with increasing Ge concentration, and further that these peaks do not form adjacent to each other but with some separation. \\

It is instructive to compare our result to previously reported single size nanoparticle embedded in alloy strategy\cite{Mingo2009}. We can calculate the thermal conductivity of this structure as a special case of the size distribution, considering nanoparticles of two sizes: atomic defects with $R$ = 0.136 nm to simulate the $Si_{0.5}Ge_{0.5}$ alloy matrix, and another larger, variable nanoparticle size with 0.8\% nanoparticle volume fraction. With the DFT dispersion, after optimizing the particle size, we obtain a thermal conductivity of 2.8 W/mK. However, with the same total Ge concentration, the minimum thermal conductivity obtained from the best distribution is 0.89 W/mK, as shown in Figure 2c. Therefore, the best distribution we have found here achieves a thermal conductivity that is more than a factor of 3 lower than that of the previous strategy.

To explain the observation of peaks in the optimal size distribution, we examine the response of the thermal conductivity as we progressively increase the mass defect concentration for a single nanoparticle size. Mathematically, we can express this quantity as $|dk/dx|$, or the change in thermal conductivity per increase in $x$. Figure 3a shows $|dk/dx|$ as function of $x$ for nanoparticles of three different radii, based on the DFT dispersion. We observe that for given nanoparticles with fixed radius, the initial addition of nanoparticles results in a large drop in thermal conductivity, but that the addition of more nanoparticles has less and less of an effect as $x$ increases.

We also examine the phonon scattering rate due to nanoparticles, $\tau_{np}^{-1}$, as a function of phonon frequency in Fig 3b. We notice that, although nanoparticles of any radius scatter phonons over the whole spectrum, each size is most effective in a limited particular bandwidth.    	  

From these two observations we can explain the discrete nature of the best distribution. Starting from a pure crystal, there exists a single nanoparticle radius, $R_1$, which can maximally lower the thermal conductivity, as quantified by $|dk/dx|$. The best choice for the first nanoparticle radius is naturally this value.  However, as in figure 3a, the thermal conductivity reduction achieved by adding more nanoparticles of this radius decreases rapidly. At a certain concentration of nanoparticles of radius $R_1$, there exists another radius, $R_2$, which can provide a larger thermal conductivity reduction than can $R_1$. Once this condition is reached, another peak around $R_2$ starts forming. 

To understand why these radii are chosen and why they do not form adjacent to each other, consider the scattering rates in Figure 3b. The first nanoparticle size is chosen to scatter the portion of the phonon spectrum that contributes maximally to thermal conductivity. However, this nanoparticle can only scatter phonons effectively in a limited bandwidth. Once the transition point to add a different radius nanoparticle is reached, the next nanoparticle radius should be chosen to scatter the portion of the spectrum that is least affected by the first nanoparticle. Correspondingly, $R_2$ should be very different from $R_1$ and chosen such that it scatters the remaining part of the phonon spectrum that is carrying heat.  These observations explain why the best size distribution is a series of non-adjacent discrete peaks.

We can also identify the best size distribution in the presence of other scattering mechanisms such as grain boundary scattering. Introducing grain boundaries into bulk materials can effectively scatter long MFP phonons, allowing nanoparticles to be used to scatter other phonons. We model this scattering mechanism using the phenomenological scattering rate $\tau_{grain}^{-1}=v/L$ , where $L$ is the average grain size. 

We present the best distribution for a grain size of 100 nm and 20 nm in Figure 4. We see that for a grain size of 20 nm and Ge concentration of 12.2\%, a thermal conductivity as low as 2.5 W/mK can be obtained with fewer peaks. This value is even smaller than the 2.6 W/mK achieved with higher Ge concentration, 17\% but without grain boundaries. This observation supports the recently introduced panoscopic concept\cite{hierarchical2012}, in which different structures are used to most effectively scatter the broad phonon spectrum. However, for higher Ge concentrations, as shown in Figures 4b and 4d, grain boundaries only result in a small additional reduction of thermal conductivity, demonstrating that mass defects alone with the best size distribution can result in exceptionally low thermal conductivity.

\section*{Discussion}
While we have identified the best size distribution that yields the minimum thermal conductivity, fabricating this distribution is still challenging due to limited experimental control over nanoparticles sizes. Therefore, we investigate a modified strategy in which we consider a distribution with point defects and a single other nanoparticle size and optimize both the size of nanoparticles and the fraction of Ge atoms in the nanoparticles. Figure 5a shows the thermal conductivities obtained from the best size distribution of this work, that of Ref. \cite{Mingo2009}, and the modified strategy at various Ge concentrations. Compared with Ref. 9, our modified strategy yields much lower thermal conductivities that are close to the values obtained from best size distribution. At a Ge concentration of ~33\%, thermal conductivity as low as that of amorphous Si, 1.5 W/mK\cite{ Cahill1994}, can be achieved. At the maximum Ge concentration of 50.8\% the modified strategy yields a thermal conductivity of 1 W/mK, still nearly a factor of 3 lower than that from Mingo et al's strategy. Further, this size distribution consists of only one nanoparticle size and thus is simpler to fabricate in practice than the best distribution. The fraction of Ge allocated to nanoparticles for the modified strategy as a function of total Ge concentration is shown in Figure 5b. At high Ge concentrations, the strategy is simple: the Ge is split approximately equally between point defects and nanoparticles with a radius of 0.816 nm.

\section*{Summary}
In conclusion, we have demonstrated that the best distribution that yields the minimum thermal conductivity of a crystal with nanoparticles is a discrete size distribution at a few well-chosen radii. Further, we demonstrated that a modified distribution with only a single nanoparticle size achieves nearly the same low thermal conductivity below that of amorphous silicon. Our work provides important insights into how to achieve the minimum thermal conductivity in crystals.
\\

\section*{Acknowledgments}
This work was supported by a start-up fund from the California Institute of Technology and by the National Science Foundation under CAREER Grant CBET 1254213.

\section*{Author contributions}
A.M. conceived this project. H.Z. designed the algorithm and performed calculations. H.Z. and A.M. analyzed the data. H.Z., A.M. discussed the result. H.Z. and A.M. wrote the manuscript. 

\section*{Additional information}
Competing financial interests: The authors declare no competing financial interests

\bibliographystyle{naturemag}

\pagebreak

\begin{figure}
  \includegraphics[width=\columnwidth]{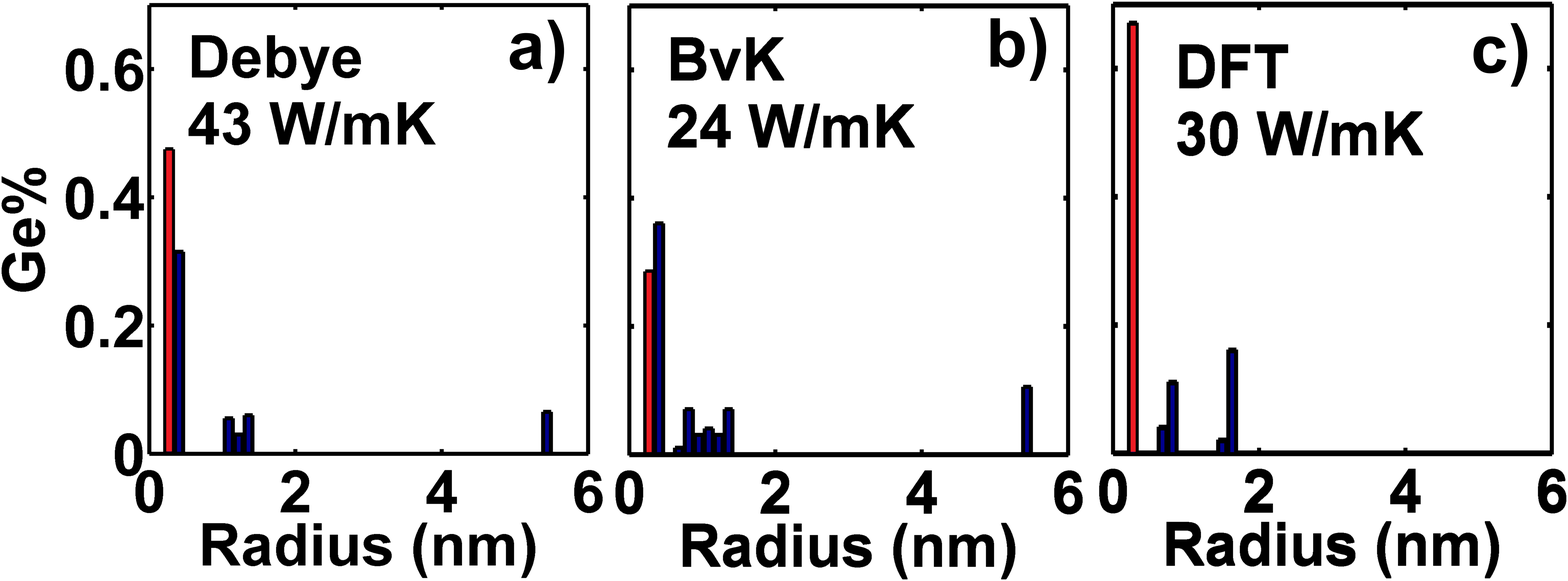}
  \caption{ Optimal nanoparticle size distribution with 1\% volumetric percentage Ge based on the (a) Debye model, (b) BvK model, and (c) DFT dispersion. The optimal distribution for each dispersion is very similar, consisting of non-adjacent discrete peaks. Red histogram bars indicate point defects.}
\end{figure}

\begin{figure}
  \includegraphics[width=\columnwidth]{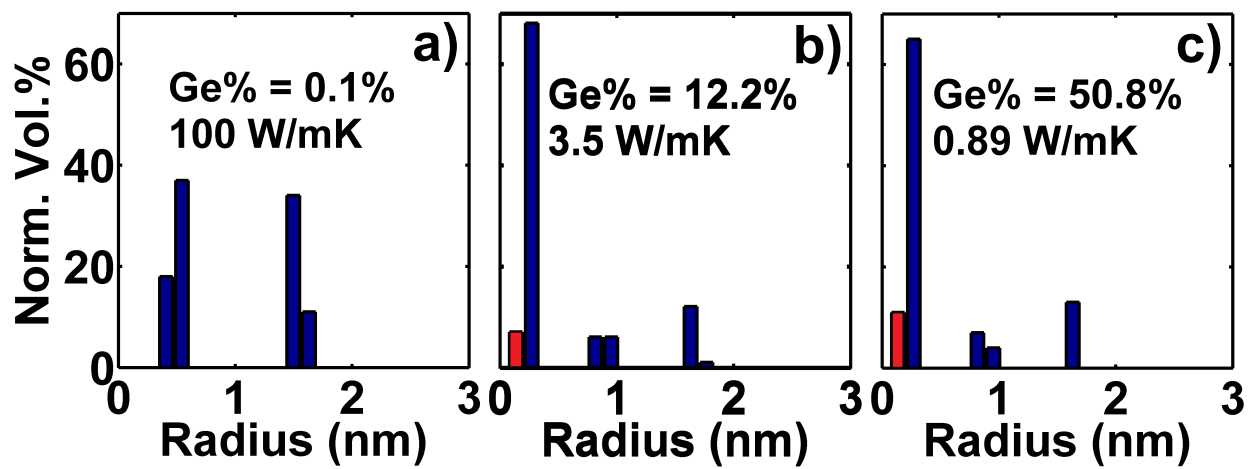}
  \caption{ Normalized optimized size distribution of nanoparticles for various Ge concentrations in bulk Si based on the DFT dispersion: for total Ge volumetric percentage of (a) Ge\% = 0.1\%, (b)Ge\% = 12.2\%, (c) Ge\% = 50.8\% . Heights of histogram bars indicate fraction of Ge allocated to each nanoparticle size. The best distribution introduces additional non-adjacent peaks as more Ge is added. Red histogram bars indicate point defects.}
\end{figure} 

\begin{figure}
  \includegraphics[width=\columnwidth]{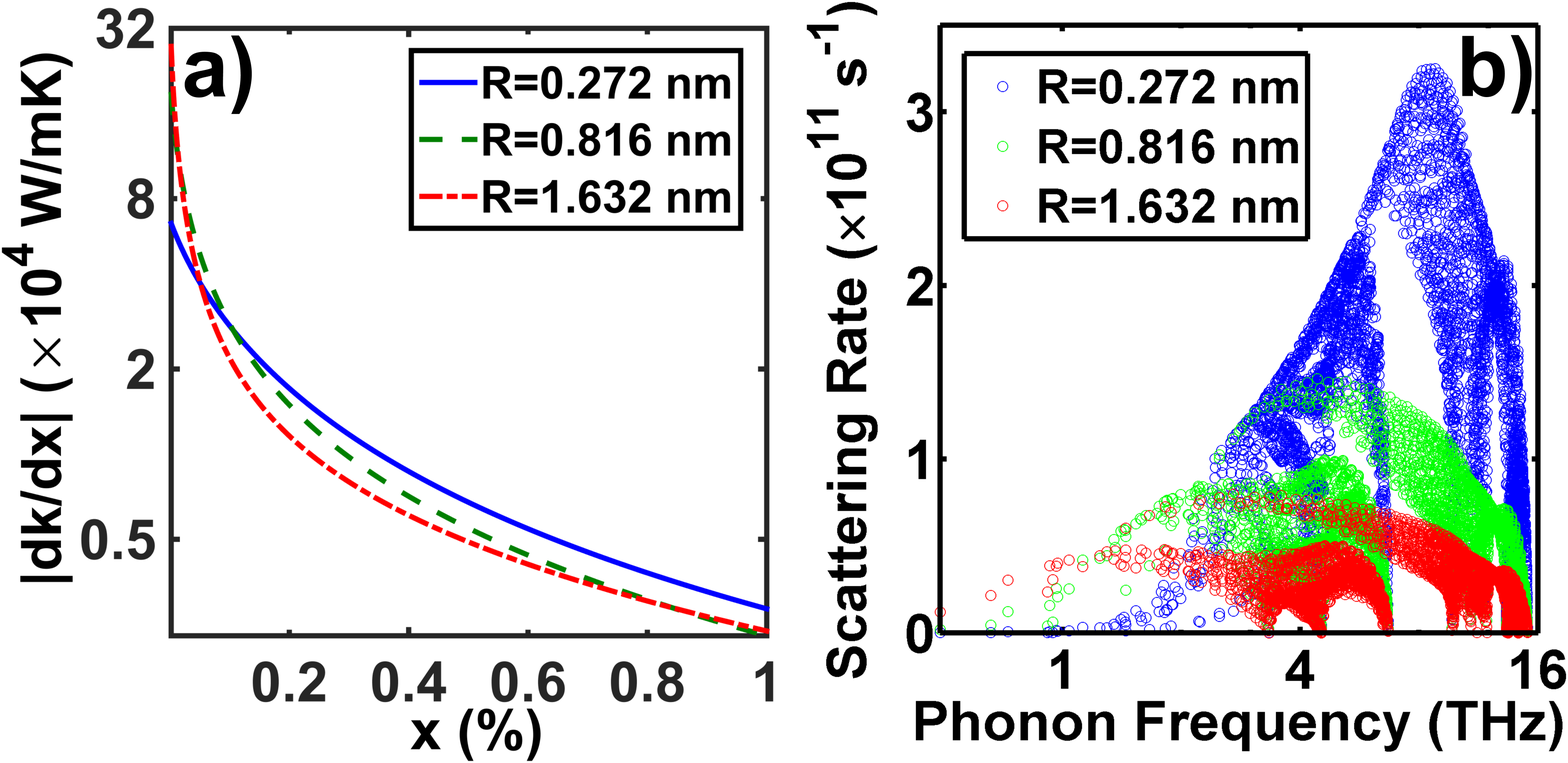}
  \caption{ (a) Reduction in thermal conductivity per change in volume fraction of Ge, $|dk/dx|$, for various particle radii ($R$ = 0.272 nm, 0.816 nm and 1.632 nm). The reduction in thermal conductivity achieved by adding more Ge defects decreases rapidly with increasing Ge concentration (b) Scattering rate of phonons by nanoparticles as a function of phonon frequency. Each circle indicates a particular phonon mode in the Brillouin zone from the DFT dispersion. Nanoparticles scatter phonons most effectively within a specific range of phonon frequencies. The Ge concentration for each radius is 1\%.}
\end{figure}

\begin{figure}
  \includegraphics[width=0.90\columnwidth]{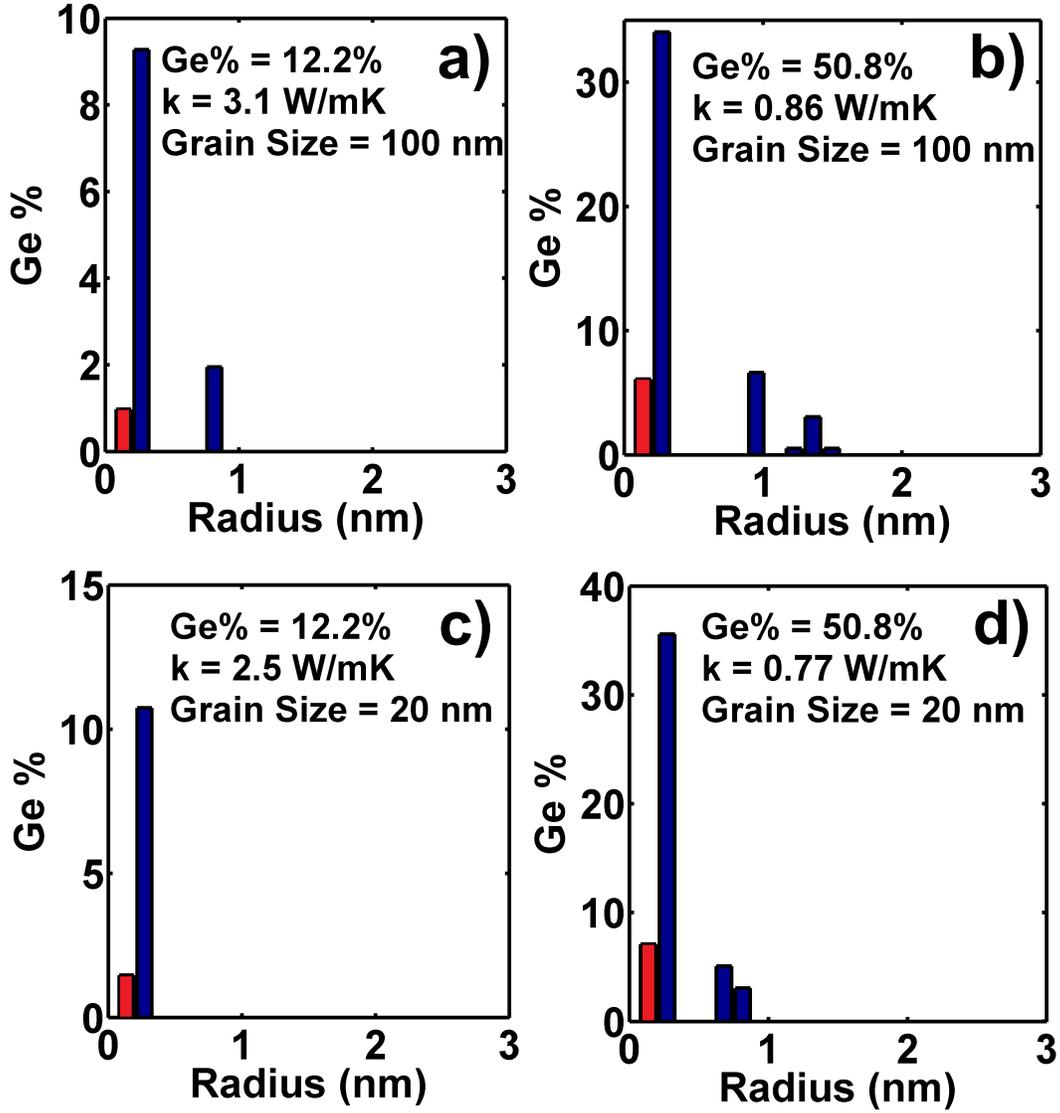}
  \caption{ Optimal size distribution of Ge nanoparticles embedded into polycrystalline Si based on DFT dispersion with: (a,b):  Averaged grain size is 100 nm with Ge concentration of (a) 12.2\% and (b) 50.8\%.(c,d): Averaged grain size is 20 nm with Ge concentration of (c) 12.2\% and (d) 50.8\%. Very low thermal conductivity can be achieved using the best distribution of Ge defects along with grain boundaries, although the thermal conductivity of 50.8\% volumetric fraction is minimally affected.}

\end{figure}

\begin{figure}
  \includegraphics[width=\columnwidth]{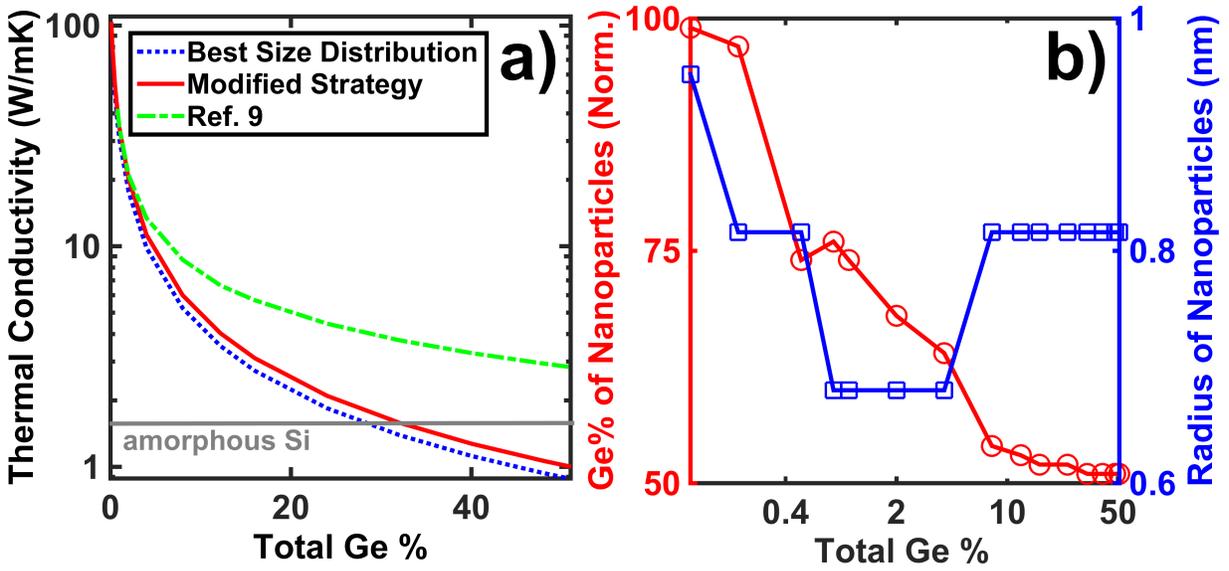}
  \caption{ (a) Thermal conductivity obtained from best size distribution, distribution of Ref. \citenum{Mingo2009} and our modified single size nanoparticle strategy as a function of total Ge concentration. The gray line indicates thermal conductivity of amorphous silicon. The modified strategy achieves nearly the same thermal conductivity as the best distribution yet with only a single nanoparticle size. (b) Radii of nanoparticles and percentage of Ge made into nanoparticles, normalized to total Ge concentration, as a function of total Ge concentration under the modified strategy. At high Ge concentrations, the optimal size of nanoparticle is a constant, and the Ge is split equally between point defects and nanoparticles.}
\end{figure}

\end{document}